\documentclass[conference]{IEEEtran}
\usepackage[utf8]{inputenc}
\usepackage[T1]{fontenc}
\usepackage{amssymb}
\usepackage{amsmath}
\usepackage{mathtools}
\usepackage{algorithm}
\usepackage{algpseudocode}
\usepackage{graphics}
\usepackage{epsfig}
\usepackage{subfigure}
\usepackage{soul,xcolor}
\usepackage{amsfonts}
\usepackage{graphicx}
\usepackage{multirow}
\usepackage{textcomp}	
\usepackage{color}
\usepackage{cite}
\usepackage[letterpaper,left=0.68in,right=0.68in,top=0.7in,bottom=1.05in]{geometry}

\IEEEoverridecommandlockouts
\allowdisplaybreaks

\begin{document}

\title{Intelligent Reflecting Surfaces and Classical Relays: Coexistence and Co-Design}

\author{\IEEEauthorblockN{Te-Yi Kan{\textsuperscript{$1$}}, Ronald Y. Chang{\textsuperscript{$1$}}, and Feng-Tsun Chien{\textsuperscript{$2$}}}
\IEEEauthorblockA{
{\textsuperscript{$1$}}Research Center for Information Technology Innovation, Academia Sinica, Taiwan \\
{\textsuperscript{$2$}}Institute of Electronics, National Yang Ming Chiao Tung University, Taiwan \\
}
\IEEEauthorblockA{Email: dexter.ty.kan@gmail.com, rchang@citi.sinica.edu.tw, ftchien@mail.nctu.edu.tw}
\thanks{This work was supported in part by the Ministry of Science and Technology, Taiwan, under Grants MOST 109-2221-E-001-013-MY3 and MOST 110-2221-E-A49-035.}}

\maketitle

\begin{abstract}
This paper investigates a multiuser downlink communication system with coexisting intelligent reflecting surface (IRS) and classical half-duplex decode-and-forward (DF) relay. In this system, the IRS and the DF relay interact with each other and assist transmission simultaneously. In particular, active beamforming at the base station (BS) and at the DF relay, and passive beamforming at the IRS, are jointly designed to maximize the sum-rate of all users. The sum-rate maximization problem is nonconvex due to the coupled beamforming vectors. We propose an alternating optimization (AO) based algorithm to tackle this complex co-design problem. Numerical validation and discussion on the superiority of the coexistence system and the tradeoffs therein are presented.
\end{abstract}

\IEEEpeerreviewmaketitle

\section{Introduction}

Intelligent reflecting surfaces (IRSs), also known as reconfigurable intelligent surfaces (RISs), passive intelligent mirrors (PIMs), and large intelligent surfaces (LISs), have recently been proposed as a key enabler for next-generation wireless communications \cite{WuZhang2020, Renzo2020}. IRSs employ a large number of low-cost passive elements, instead of active transmitters, to ``reflect'' signals to empower smart and reconfigurable radio environments. It is envisioned that deploying IRSs is less expensive as compared to installing active transmitters such as base stations (BSs) and relays, and operating IRSs is more energy-efficient and eases signal processing and interference management requirements due to its low-cost and passive nature. Thus, IRSs are suitable for dense deployment and are promising to help meet the demands of beyond fifth-generation (B5G) and sixth-generation (6G) communications, such as high data rates, energy efficiency, and ubiquitous connectivity.

There is a wealth of research on IRS focusing on joint transmit beamforming and IRS phase shifts optimization for sum-rate maximization (e.g., \cite{Wu2018, Huang2018, ML_Huang2020, OFDM_Li2020, Mu2020}) or power minimization (e.g., \cite{Wu2019, NOMA_Zheng2020}). IRS-enhanced multi-input single-output (MISO) systems were studied for single-user \cite{Wu2018} and multiuser \cite{Huang2018} scenarios, where transmit beamforming at the BS and the phase shifts at the IRS were configured by optimizing the received signal power or sum-rate of users. Machine learning-based techniques were also introduced into IRS-aided systems to adjust transmit beamforming and IRS phase shifts \cite{ML_Huang2020}.
IRS-enhanced broadband orthogonal frequency division multiplexing (OFDM) systems were also studied \cite{OFDM_Li2020}.
In \cite{OFDM_Li2020}, BS beamforming and IRS reflection were jointly designed for the objective of sum-rate maximization or, after transformation, mean square error (MSE) minimization in multiuser MISO systems. IRS-empowered non-orthogonal multiple access (NOMA) systems for signal-to-interference-plus-noise ratio (SINR) improvement were also studied \cite{Mu2020}. On the other hand, power minimization problems in IRS-aided systems were considered \cite{Wu2019, NOMA_Zheng2020}. In \cite{Wu2019}, transmit beamforming and IRS beamforming for power minimization subject to user SINR constraints in MISO systems were considered. In \cite{NOMA_Zheng2020}, transmit power minimization in IRS-empowered NOMA systems was studied. It was shown that incorporating IRSs into NOMA brings merits in decoding and differentiating multiple users due to IRS's ability to reconfigure wireless channels.

Similarities and differences between IRS and classical decode-and-forward (DF) and amplify-and-forward (AF) relaying have been explored \cite{Bjornson2020, Huang2019, Nadeem2020, Renzo2020_relay}. In \cite{Bjornson2020}, comparisons between IRS and DF relaying were conducted analytically and numerically for a SISO system, showing that IRS with sufficient reflecting elements can achieve higher data rates and better energy efficiency in high target data rate regimes than DF relaying. In \cite{Huang2019}, comparisons between IRS and AF relaying were conducted numerically in an IRS-aided multiuser MISO system. It was shown that replacing the IRS with an AF relay yields higher sum rates but lower energy efficiency due to active power amplification of the AF relay. In \cite{Nadeem2020}, IRS and full-duplex/half-duplex AF relaying were compared. It was shown that IRS achieves comparable or even better performance as compared to full-duplex and half-duplex AF relaying when the number of IRS reflecting elements is sufficiently large. In \cite{Renzo2020_relay}, a comprehensive discussion on the subject was presented from various perspectives such as hardware complexity, power consumption, and spectral efficiency. It was similarly concluded that IRS with sufficient reflecting elements has the potential to outperform relay-aided transmission since IRS operates in a full-duplex manner, yet without the loop interference as in full-duplex relays.

In this paper, we consider the practical scenario with {\it coexisting} IRS and half-duplex DF relay in a multiuser MISO (MU-MISO) system. The consideration is interesting and new, in especially two aspects. First, the IRS and relay coexistence system is different from the traditional multi-relay (in this case, two-relay) system, since there was no interaction between the relays in a multi-relay system but there is interaction between IRS and relay in the new system. Second, the IRS reflects signals at both phases of half-duplex DF relaying, and there exist tradeoffs in the design of IRS beamforming to cater for both the end users and the DF relay. The complex co-design problem of BS beamforming, relay beamforming, and IRS beamforming for maximum sum rates is examined. Note that these aspects were unexplored in the recent related work \cite{Yildirim2021}. Finally, numerical validation and discussion on the superiority of the coexistence system and tradeoffs therein are presented.

\section{System Model and Problem Formulation} \label{sec_system_problem}
\subsection{System Model} \label{subsec_system}
\begin{figure}[tb!]
    \centering
    \includegraphics[width={0.83\linewidth}]{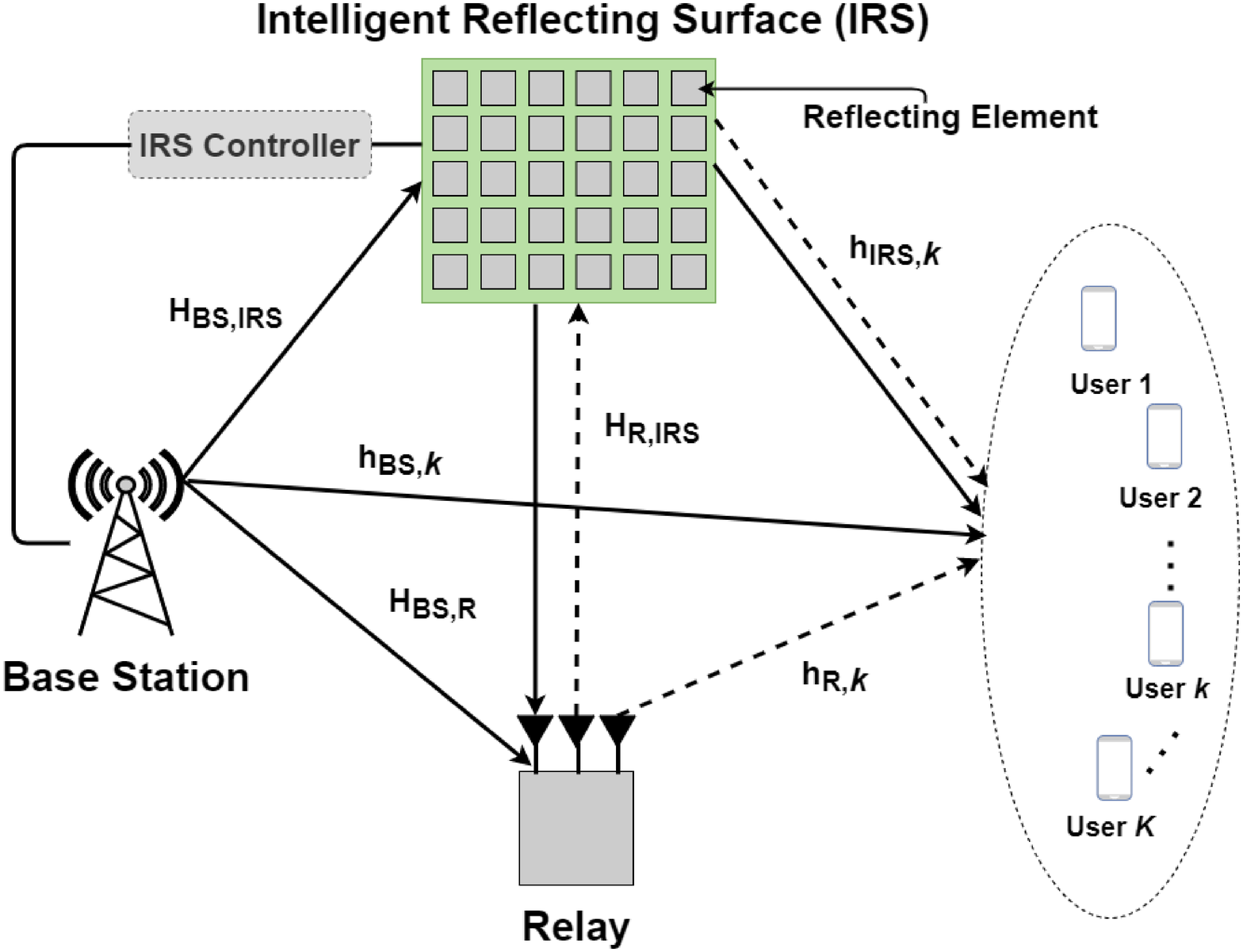}
    \caption{A coexisting IRS and relay assisted MU-MISO system. Solid/dashed lines indicate transmissions in the first/second phases.}
    \label{fig:system_model}
\vspace{-0.1in}
\end{figure}
As shown in Fig.~\ref{fig:system_model}, we consider a downlink MU-MISO communication system that comprises a BS with $M$ antennas, $K$ single-antenna end users, an IRS with $N$ reflecting elements, and a classical half-duplex DF relay with $L$ antennas. Here, to ensure a sufficient degree of freedom, we assume $K \leq \min \{M, L\}$.  The baseband equivalent channels between two communication nodes among the BS, IRS, relay, and user $k$ are denoted by self-explanatory notations ${\mathbf H}_{{\rm BS}, {\rm R}} \in {\mathbb C}^{L \times M}$, ${\mathbf H}_{{\rm BS}, {\rm IRS}} \in {\mathbb C}^{N \times M}$, ${\mathbf h}_{{\rm BS},k} \in {\mathbb C}^{1 \times M}$,  ${\mathbf H}_{{\rm R}, {\rm IRS}} \in {\mathbb C}^{N \times L}$, ${\mathbf h}_{{\rm R},k} \in {\mathbb C}^{1 \times L}$, and ${\mathbf h}_{{\rm IRS},k} \in {\mathbb C}^{1 \times N}$.  All channels are assumed to be quasi-static and Rayleigh flat-fading. The small-scale fading is modeled by complex Gaussian with zero mean and unit variance and the large-scale fading is modeled by
$
\kappa {\left({d}/{d_{0}}\right)}^{-\varrho},
$
where $d$ is the distance between the two end nodes, $d_{0}$ is the reference distance, $\kappa$ is a large-scale fading constant, and $\varrho$ is the path-loss exponent. The CSI of all channels is assumed perfectly known at the BS, the IRS controller, and the relay. Due to the presence of the relay, there are two phases in one complete transmission of information, for which the first phase and the second phase are assumed to be within the channel coherence interval. Moreover, in this paper, reflecting elements can be adjusted by the IRS controller once in each complete transmission of the information. Next, we elaborate on the two-phase transmission protocol.

\subsubsection{First Phase}
During the first phase, the BS transmits its signal to all users and the relay via direct link. Meanwhile, the IRS reflects the incident signal from the BS towards the DF relay and all users. The received signal at user $k$ in the first phase is given by
\begin{align} \label{eq_received_signal_I}
{y}_{k}^{\rm I} ={\mathbf h}^{\prime}_{{\rm BS},k} {\mathbf x} + w_{k}^{\rm I},
\end{align}
where ${\mathbf h}^{\prime}_{{\rm BS},k} \triangleq  {\mathbf h}_{{\rm IRS},k} {\mathbf \Theta} {\mathbf H}_{{\rm BS}, {\rm IRS}} + {\mathbf h}_{{\rm BS},k}$ is the effective channel from the BS to the end user $k$, ${\mathbf \Theta} \triangleq {\rm diag}({\boldsymbol \theta})$ with ${\boldsymbol \theta} = \begin{bmatrix} \theta_{1} & \ldots & \theta_{N} \end{bmatrix}$ is the diagonal matrix accounting for the passive IRS beamforming, and $w_{k}^{\rm I} \sim \mathcal {CN} \left( 0, \sigma_{k}^2 \right)$ denotes the additive white Gaussian noise (AWGN) at user $k$  during the first phase. Note that the $n$-th reflecting element adjustment $\theta_{n}$  is given by $\theta_{n} = \beta_{n} e^{j \psi_{n}}$ with the amplitude attenuation $\beta_{n} \leq 1$ and the phase shift $\psi_{n} \in \left[0,2\pi\right)$.

The transmit signal at the BS is ${\mathbf x}= \sum_{k=1}^{K} {\mathbf g}_{k}s_{k} = {\mathbf G} \cdot {\mathbf s}$, where ${\mathbf g}_{k} \in {\mathbb C}^{M \times 1}$ denotes the beamforming vector for user $k$ with ${\mathbf G} \triangleq [{\mathbf g}_{1}, \ldots, {\mathbf g}_{K}]$ and ${\mathbf s} \triangleq [s_{1}, \ldots, s_{K}]^{\rm T}$, $s_{k}$ is the data intended for user $k$ with zero mean and unit variance. The data streams from different users are assumed independent. Then, the transmit power at the BS is $P_{\rm BS}^{\rm total} =  {\mathbb E}\big[{\Vert {\mathbf x} \Vert}^2\big] = {\rm tr}\left({\mathbf G}{\mathbf G}^{\rm H} \right) \leq P^{\max}_{\rm BS},$
where $P^{\max}_{\rm BS}$ is the maximum available power at the BS. The SINR for user $k$ in the first phase is 
\begin{align} \label{eq_SINR_I}
\gamma_{k}^{\rm I} = \frac{{ |{\mathbf h}^{\prime}_{{\rm BS},k} {\mathbf g}_{k} |}^2}{\sum_{j=1, j \neq k}^{K} { |{\mathbf h}^{\prime}_{{\rm BS},k} {\mathbf g}_{j} |}^2 +  \sigma_{k}^2}.
\end{align}
On the other hand, the received signal at the DF relay is
\begin{align} \label{eq_received_Relay}
{\mathbf y}_{\rm R} = {\mathbf H}^{\prime}_{{\rm BS}, {\rm R}} {\mathbf x} + {\mathbf w}_{\rm R},
\end{align}
where ${\mathbf H}^{\prime}_{{\rm BS}, {\rm R}} \triangleq {\mathbf H}_{{\rm R}, {\rm IRS}}^{\rm H} {\mathbf \Theta} {\mathbf H}_{{\rm BS}, {\rm IRS}} + {\mathbf H}_{{\rm BS}, {\rm R}}$ is the effective channel from the BS to the relay and ${\mathbf w}_{\rm R} \sim \mathcal{CN} \left(0, \sigma_{R}^2 {\mathbf I}_{L} \right)$ is the AWGN at the relay. Matched filtering is employed at the relay to decode all users' signals from the BS. The resulting SINR is \cite{Tse2005}
\begin{align} \label{eq_SINR_relay}
\gamma_{{\rm R},k}
= \frac{{\Vert {\boldsymbol \alpha}_{k} \Vert}^{4}}{\sum_{j=1, j \neq k}^{K} {\boldsymbol \alpha}_{k}^{\rm H} {\boldsymbol \alpha}_{j} {\boldsymbol \alpha}_{j}^{\rm H} {\boldsymbol \alpha}_{k} + \sigma_{\rm R}^{2} {\Vert {\boldsymbol \alpha}_{k} \Vert}^{2}},
\end{align}
where  $ {\boldsymbol \alpha}_{k} \triangleq {\mathbf H}^{\prime}_{{\rm BS}, {\rm R}} {\mathbf g}_{k}$ is the filter weight for user $k$.

\subsubsection{Second phase}
During the second phase, the relay transmits the signal to all users and the IRS, where the latter also reflects the incident signal towards all users. Here, we assume that the DF relay can perfectly decode the signal for user $k$ if the SINR $\gamma_{{\rm R}, k}$ for user $k$ at the relay exceeds a predefined threshold $\gamma_{\rm R}^{\rm th}$. 
The transmit signal at the relay is
$
{\mathbf x}_{\rm R} = \sum_{k=1}^{K} {\mathbf f}_{k} s_{k} = {\mathbf F} \cdot {\mathbf s},
$
where ${\mathbf f}_{k} \in {\mathbb C}^{L \times 1}$ denotes the beamforming vector at the relay for user $k$ and ${\mathbf F} \triangleq [{\mathbf f}_{1}, \ldots, {\mathbf f}_{K}]$. The relay's transmit power is $P_{\rm R}^{\rm total} =  {\mathbb E}\left[{\Vert {\mathbf x}_{\rm R} \Vert}^2\right] = {\rm tr}\left({\mathbf F} {\mathbf F}^{\rm H}\right) \leq P_{\rm R}^{\max},$
where $P_{\rm R}^{\max}$ denotes the relay's maximum available power. The received signal at user $k$ in the second phase can be represented by
\begin{align} \label{eq_received_signal_II_DF}
y_{k}^{\rm II} =  {\mathbf h}^{\prime}_{{\rm R},k} {\mathbf x}_{\rm R} + w_{k}^{\rm II},
\end{align}
where ${\mathbf h}^{\prime}_{{\rm R},k} \triangleq {\mathbf h}_{{\rm IRS},k} {\mathbf \Theta} {\mathbf H}_{{\rm R}, {\rm IRS}} + {\mathbf h}_{{\rm R},k}$ is the effective channel from the relay to user $k$ and  $w_{k}^{\rm II} \sim \mathcal{CN} \left( 0, \sigma_{k}^2 \right)$ is the corresponding AWGN. It follows that the SINR for user $k$ in the second phase is
\begin{align} \label{eq_SINR_II}
\gamma_{k}^{\rm II} = \frac{{ |{\mathbf h}^{\prime}_{{\rm R}, k} {\mathbf f}_{k} |}^2}{\sum_{j=1, j \neq k}^{K} { |{\mathbf h}^{\prime}_{{\rm R}, k} {\mathbf f}_{j} |}^2 +  \sigma_{k}^2}.
\end{align}
Finally, the resulting  SINR after combining the signals from the two phases using the maximal ratio combining is
\begin{align} \label{eq_SINR}
\gamma_{k} = \gamma_{k}^{\rm I} + \gamma_{k}^{\rm II}.
\end{align}
%

\subsection{Problem Formulation} \label{subsec_problem}
In this paper, we aim for maximizing the sum-rate of all the users by jointly optimizing the active beamforming at the BS, the active beamforming at the relay, and the passive beamforming at the IRS. Specifically, the optimization problem is formulated as
\begin{subequations} \label{eq_op_sum_rate}
\begin{align}
& \max_{\{\mathbf{g}_{k}\}, \{\mathbf{f}_{k}\}, \mathbf{\Theta}}
& & \sum_{k=1}^{K}{\log_{2}\left(1 + \gamma_{k}\right)} \label{eq_op_sum_rate_obj}  \\
& \hspace{20pt} \text{s.t.} & & P_{\rm BS}^{\rm total} \leq P^{\max}_{\rm BS}, \label{eq_op_sum_rate_cons_1} \\
& & & P_{\rm R}^{\rm total} \leq P_{\rm R}^{\max}, \label{eq_op_sum_rate_cons_2}  \\
& & & {\left|\theta_{n}\right|} \leq 1, \forall n = 1,2,\ldots,N , \label{eq_op_sum_rate_cons_3}  \\
& & & \gamma_{{\rm R},k} \geq \gamma_{\rm R}^{\rm th}, \forall k = 1,2, \ldots,K, \label{eq_op_sum_rate_cons_4}
\end{align}
\end{subequations}
where the optimization variables and objective function are given in \eqref{eq_op_sum_rate_obj}, $\{\mathbf{g}_{k}\}$ and $\{\mathbf{f}_{k}\}$ are the sets of the BS beamforming and the relay beamforming vectors for all users, respectively. Constraints \eqref{eq_op_sum_rate_cons_1} and \eqref{eq_op_sum_rate_cons_2} denote the available transmit power at the BS and the DF relay, respectively. Constraint \eqref{eq_op_sum_rate_cons_3} is the reflection constraint for an ideal IRS. Constraint \eqref{eq_op_sum_rate_cons_4} ensures the decoding SINR requirements for all users' signals at the relay are satisfied. Note that the pre-log factor of $1/2$ due to two-phase transmission is dropped in the achievable rate expression, but is considered in all simulations.

Solving the problem in \eqref{eq_op_sum_rate} is a challenging task, as the optimization variables are highly coupled, rendering a nonconvex problem. To tackle the problem, we propose an alternating optimization (AO)-based algorithm to decouple \eqref{eq_op_sum_rate} into three subproblems, as detailed in the next section.

\section{AO-Based Beamforming Co-Design} \label{sec_pro_alg}

In this section, we elaborate on the three subproblems in the proposed AO-based algorithm: BS beamforming, relay beamforming, and IRS beamforming. The entire AO algorithm runs through optimizing the BS beamforming (in Sec. \ref{sec_pro_alg_BS}), relay beamforming (in Sec. \ref{sec_pro_alg_Relay}), and IRS beamforming (in Sec. \ref{sec_pro_alg_IRS}), in an alternate fashion.

\subsection{BS Beamforming Optimization} \label{sec_pro_alg_BS}

Given fixed relay beamforming vectors $\{{\mathbf f}_{k}\}$ and IRS beamforming matrix $\mathbf \Theta$, the problem in \eqref{eq_op_sum_rate} becomes
\begin{subequations} \label{eq_op_AO_BS}
\begin{align}
& \max_{\{\mathbf{g}_{k}\}}
& & \sum_{k=1}^{K}{\log_{2}\left(C_{1,k} + \gamma_{k}^{\rm I}\right)} \label{eq_op_AO_BS_obj}  \\
& \hspace{9pt} \text{s.t.} & & P_{\rm BS}^{\rm total} \leq P^{\max}_{\rm BS}, \label{eq_op_AO_BS_cons1} \\
& & & \gamma_{{\rm R},k} \geq \gamma_{\rm R}^{\rm th},  k = 1,\ldots,K, \label{eq_op_AO_BS_cons2}
\end{align}
\end{subequations}
where $C_{1,k} = 1 + \gamma_{k}^{\rm II},$  $k = 1,\ldots,K$ are considered as constants under the AO-based decomposition. The subproblem \eqref{eq_op_AO_BS} is still nonconvex and difficult to solve. In this subsection, we tackle the subproblem in \eqref{eq_op_AO_BS}  by first transforming the problem of finding $\mathbf{g}_{k}$ into one of finding the rank-one positive semidefinite matrix ${\mathbf G}_{k}=\mathbf{g}_{k}{\mathbf g}_{k}^{\rm H}$ and then by relaxing the rank-one constraint, which eventually leads to a convex semidefinite programming (SDP) problem. Toward this end, we develop the following four steps 1) introducing slack variables, 2) first-order Taylor approximation, 3) alternating optimization, and 4) semidefinite relaxation, as detailed below.

\subsubsection{Introducing Slack Variables}
We introduce slack variables $\{{\mathcal S}_{1,k}\}$, $\{{\mathcal I}_{1,k}\}$, $\{{\mathcal S}_{{\rm R},k}\}$, and $\{{\mathcal I}_{{\rm R},k}\}$ that satisfy
\begin{align}
\frac{1}{{\mathcal S}_{1,k}} &\triangleq {\left |{\mathbf h}^{\prime}_{{\rm BS},k} {\mathbf g}_{k} \right|}^2 \leq {\rm tr}({\mathbf G}_{k} {\mathbf h}^{\prime \prime}_{{\rm B},k}), \label{S_1}\\
{\mathcal I}_{1,k} & \triangleq \sum_{j \neq k} {\left |{\mathbf h}^{\prime}_{{\rm BS},k} {\mathbf g}_{j} \right|}^2 + \sigma_{k}^{2} \geq \sum_{j \neq k}{\rm tr}({\mathbf G}_{j} {\mathbf h}^{\prime \prime}_{{\rm B},k}) + \sigma_{k}^{2},\label{I_1}\\%
\frac{1}{{\mathcal S}_{{\rm R},k}} &\triangleq {\Vert {\boldsymbol \alpha}_{k} \Vert}^{4} \leq {[{\rm tr}({\mathbf G}_{k} {\mathbf H}_{{\rm B}, {\rm R}}^{\prime \prime})]}^{2},\label{S_R}\\%
{\mathcal I}_{{\rm R},k} &\triangleq \sum_{j\neq k} {\boldsymbol \alpha}_{k}^{\rm H} {\boldsymbol \alpha}_{j} {\boldsymbol \alpha}_{j}^{\rm H} {\boldsymbol \alpha}_{k} + \sigma_{\rm R}^{2} {\Vert {\boldsymbol \alpha}_{k} \Vert}^{2} \nonumber \\
&\geq \sum_{j\neq k}{\rm tr}({\mathbf G}_{k} {\mathbf H}_{{\rm B}, {\rm R}}^{\prime \prime} {\mathbf G}_{j} {\mathbf H}_{{\rm B}, {\rm R}}^{\prime \prime}) + \sigma_{\rm R}^{2} {\rm tr}({\mathbf G}_{k} {\mathbf H}_{{\rm B}, {\rm R}}^{\prime \prime}),\label{I_R} 
\end{align}
where ${\mathbf H}_{{\rm B}, {\rm R}}^{\prime \prime} \triangleq {{\mathbf H}_{{\rm BS}, {\rm R}}^{\prime}}^{\rm H} {\mathbf H}_{{\rm BS}, {\rm R}}^{\prime}$ and ${\mathbf h}^{\prime \prime}_{{\rm B},k} \triangleq {{\mathbf h}^{\prime}_{{\rm BS},k}}^{\rm H} {\mathbf h}^{\prime}_{{\rm BS},k}$.
Incorporating these slack variables into \eqref{eq_op_AO_BS} gives rise to the following equivalent problem
\begin{subequations} \label{eq_op_AO_BS_2}
\begin{align}
& \max
& & \sum_{k=1}^{K}{R_{1,k}} \label{eq_op_AO_BS_2_obj}  \\
& \text{variables: } & & \{\mathbf{G}_{k}\}, \{R_{1,k}\}, \{{\mathcal S}_{1,k}\}, \{{\mathcal I}_{1,k}\}, \{{\mathcal S}_{{\rm R},k}\}, \{{\mathcal I}_{{\rm R},k}\} \nonumber  \\
& \hspace{5pt} \text{s.t.} & & R_{1,k} \leq \log_{2}\big(C_{k,1} + \frac{1}{{\mathcal S}_{1,k}{{\mathcal I}_{1,k}}}\big)\triangleq u_1({\bf z}), \label{eq_op_AO_BS_2_cons1} \\
& & &  v_1({\bf z}_{\rm R})\triangleq\frac{1}{{\mathcal S}_{{\rm R},k}{\mathcal I}_{{\rm R},k}} \geq \gamma_{\rm R}^{\rm th},\label{eq_op_AO_BS_2_cons3} \\
& & & \eqref{eq_op_AO_BS_cons1},\eqref{S_1},\eqref{I_1},\eqref{S_R},\eqref{I_R}, \label{eq_op_AO_BS_2_cons_supp} \\
& & & {\mathbf G}_{k} \succeq 0,~~ \text{rank}\left({\mathbf G}_{k}\right) = 1, \label{eq_op_AO_BS_2_cons8}\\ 
& & & \quad\quad \forall k = 1,2, \ldots,K,\notag
\end{align}
\end{subequations}
where $u_1({\bf z})$ and $v_1({\bf z}_{\rm R})$ with ${\bf z}=[{\mathcal S}_{1,k}, {\mathcal I}_{1,k}]^{\rm T}$ and ${\bf z}_{\rm R}=[{\mathcal S}_{{\rm R}, k}, {\mathcal I}_{{\rm R}, k}]^{\rm T}$ defined in \eqref{eq_op_AO_BS_2_cons1} and \eqref{eq_op_AO_BS_2_cons3}, respectively, are for notational convenience and clarity. 
Note that the optimal solution of \eqref{eq_op_AO_BS_2} meets the constraints \eqref{S_1}--\eqref{I_R} in \eqref{eq_op_AO_BS_2_cons_supp} with equality. The subproblem in \eqref{eq_op_AO_BS_2} is still nonconvex. 

\subsubsection{First-Order Taylor Approximation}
To convexify the constraints \eqref{eq_op_AO_BS_2_cons1} and \eqref{eq_op_AO_BS_2_cons3}, we utilize the fact that any convex function can be lower bounded by its first-order Taylor approximation. Since $u_1({\bf z})$ and $v_1({\bf z}_{\rm R})$ are convex, from the first-order Taylor representation, we have
\begin{align}
u_1({\bf z})&\geq u_1({\bf z}^{\rm loc})+\nabla u_1({\bf z}^{\rm loc})^{\rm T}({\bf z}-{\bf z}^{\rm loc})\triangleq R_{1, k}^{\rm low}, \label{lower bound: eq_op_AO_BS_2_cons1}\\
v_1({\bf z}_{\rm R})&\geq v_1({\bf z}_{\rm R}^{\rm loc})+\nabla v_1({\bf z}_{\rm R}^{\rm loc})^{\rm T}({\bf z}_{\rm R}-{\bf z}_{\rm R}^{\rm loc})\triangleq  \gamma_{{\rm R}, k}^{\rm low},\label{lower bound: eq_op_AO_BS_2_cons3}
\end{align}
where ${\bf z}^{\rm loc}=[{\mathcal S}_{1,k}^{\rm loc}, {\mathcal I}_{1,k}^{\rm loc}]^{\rm T}$ and ${\bf z}_{\rm R}^{\rm loc}=[{\mathcal S}_{{\rm R},k}^{\rm loc}, {\mathcal I}_{{\rm R},k}^{\rm loc}]^{\rm T}$ are local points at which $u_1({\bf z})$ and $v_1({\bf z}_{\rm R})$ are differentiable, respectively. The two lower bounds $R_{1, k}^{\rm low}$ and $\gamma_{{\rm R}, k}^{\rm low}$ can be carried out in closed-form from the gradients $\nabla u_1({\bf z}^{\rm loc})$ and $\nabla v_1({\bf z}_{\rm R}^{\rm loc})$, respectively.
Replacing $u_1({\bf z})$ in \eqref{eq_op_AO_BS_2_cons1} and $v_1({\bf z}_{\rm R})$ in \eqref{eq_op_AO_BS_2_cons3} with their respective lower bounds in \eqref{lower bound: eq_op_AO_BS_2_cons1} and \eqref{lower bound: eq_op_AO_BS_2_cons3} linearizes and convexifies the constraints \eqref{eq_op_AO_BS_2_cons1} and \eqref{eq_op_AO_BS_2_cons3}, converting the problem in \eqref{eq_op_AO_BS_2} to
\begin{subequations} \label{eq_op_AO_BS_3}
\begin{align}
& \max
& & \sum_{k=1}^{K}{R_{1,k}} \label{eq_op_AO_BS_3_obj}  \\
& \text{variables: } & & \{\mathbf{G}_{k}\}, \{R_{1,k}\}, \{{\mathcal S}_{1,k}\}, \{{\mathcal I}_{1,k}\}, \{{\mathcal S}_{{\rm R},k}\}, \{{\mathcal I}_{{\rm R},k}\} \nonumber  \\
& \hspace{9pt} \text{s.t.} & & R_{1,k} \leq R_{1, k}^{\rm low},~\gamma_{{\rm R}, k}^{\rm low} \geq \gamma_{\rm R}^{\rm th}, ~\eqref{eq_op_AO_BS_2_cons_supp}, \eqref{eq_op_AO_BS_2_cons8}, \label{eq_op_AO_BS_3_cons1} \\
& & & \quad\quad\forall k = 1,2, \ldots,K.\nonumber
\end{align}
\end{subequations}
Note that the solution set of \eqref{eq_op_AO_BS_3} is a subset of that of the original problem in \eqref{eq_op_AO_BS_2}, as lower bounds are used as substitutes in the constraints \eqref{eq_op_AO_BS_2_cons1} and \eqref{eq_op_AO_BS_2_cons3}.

\subsubsection{Alternating Optimization} The interleaving presence of BS beamforming matrices ${\mathbf G}_{k}$ and ${\mathbf G}_{j}$ still render the constraint \eqref{I_R} in \eqref{eq_op_AO_BS_2_cons_supp} nonconvex. AO is adopted within the subproblem to solve ${\mathbf G}_{k}$, $k=1,\ldots,K$, iteratively. Specifically, in the $k'$-th iteration, only the $k'$-th beamforming matrix ${\mathbf G}_{k'}$ is solved while keeping other ${\mathbf G}_{j}$, $j \neq k'$ as constants. As such, constraint \eqref{I_R} in \eqref{eq_op_AO_BS_2_cons_supp} becomes convex.

\subsubsection{Semidefinite Relaxation}
Finally, we relax the rank-one constraint in \eqref{eq_op_AO_BS_2_cons8} which, after combining with the Taylor approximation and alternating optimization, transforms the entire optimization to an SDP problem that can be solved by standard convex optimization software such as CVX \cite{cvx}. In general, the optimal solutions for ${\mathbf G}_{k}$'s are not necessarily of rank-one. One typical approach to obtaining rank-one solutions is  by using the randomization procedure \cite{Huang2014} based on the eigen-decomposition of the optimal ${\mathbf G}_{k}$.

\subsection{Relay Beamforming Optimization} \label{sec_pro_alg_Relay}

Given fixed $\{{\mathbf g}_{k}\}$ (obtained from the steps in Sec. \ref{sec_pro_alg_BS}) and $\mathbf \Theta$, the problem of solving the relay beamforming  $\{\mathbf{f}_{k}\}$ in \eqref{eq_op_sum_rate} becomes
\begin{subequations} \label{eq_op_AO_Relay}
\begin{align}
& \max_{\{\mathbf{f}_{k}\}}
& & \sum_{k=1}^{K}{\log_{2}\left(C_{2,k} + \gamma_{k}^{\rm II}\right)} \label{eq_op_AO_Relay_obj}  \\
& \hspace{9pt} \text{s.t.} & & P_{\rm R}^{\rm total} \leq P_{\rm R}^{\max}, \label{eq_op_AO_Relay_cons1}
\end{align}
\end{subequations}
where $C_{2,k} = 1 + \gamma_{k}^{\rm I},  k=1,2,\ldots,K$, are kept as constants. The subproblem for finding the relay beamforming vector $\mathbf{f}_{k}$ in \eqref{eq_op_AO_Relay} is nonconvex. Following similar steps developed in Sec.~\ref{sec_pro_alg_BS}, we elaborate on how to convexify \eqref{eq_op_AO_Relay} below.

Let ${\mathbf F}_{k} = {\mathbf f}_{k} {\mathbf f}_{k}^{\rm H}$.  Define the slack variables $\{{\mathcal S}_{2,k}\}$ and $\{{\mathcal I}_{2,k}\}$ that satisfy
\begin{align}
\frac{1}{{\mathcal S}_{2,k}} &\triangleq {\left |{\mathbf h}^{\prime}_{{\rm R},k} {\mathbf f}_{k} \right|}^2 \leq {\rm tr}\left({\mathbf F}_{k} {\mathbf h}^{\prime \prime}_{{\rm R},k}\right),\label{S_2}\\
{\mathcal I}_{2,k} & \triangleq \sum_{j \neq k} {\left |{\mathbf h}^{\prime}_{{\rm R},k} {\mathbf f}_{j} \right|}^2 + \sigma_{k}^{2} \geq \sum_{j \neq k}{\rm tr}\left({\mathbf F}_{j} {\mathbf h}^{\prime \prime}_{{\rm R},k}\right) + \sigma_{k}^{2},\label{I_2}
\end{align}
where ${\mathbf h}^{\prime \prime}_{{\rm R},k} \triangleq {{\mathbf h}^{\prime}_{{\rm R},k}}^{\rm H} {\mathbf h}^{\prime}_{{\rm R},k}$. With the above slack variables, the subproblem \eqref{eq_op_AO_Relay} can be equivalently rewritten as
\begin{subequations} \label{eq_op_AO_Relay_2}
\begin{align}
& \max
& & \sum_{k=1}^{K}{R_{2,k}} \label{eq_op_AO_Relay_2_obj}  \\
& \text{variables: } & & \{\mathbf{F}_{k}\}, \{R_{2,k}\}, \{{\mathcal S}_{2,k}\}, \{{\mathcal I}_{2,k}\} \nonumber  \\
& \hspace{9pt} \text{s.t.} & & R_{2, k} \leq \log_{2}\big(C_{2,k} + \frac{1}{{\mathcal S}_{2,k} {\mathcal I}_{2,k}}\big)\triangleq u_2({\bf z}), \label{eq_op_AO_Relay_2_cons1} \\
& & & \eqref{eq_op_AO_Relay_cons1},\eqref{S_2},\eqref{I_2}, ~{\mathbf F}_{k} \succeq 0, ~\text{rank}({\mathbf F}_{k}) = 1, \label{eq_op_AO_Relay_2_cons_supp2}\\
& & &\quad\quad\forall k = 1,2, \ldots,K,\nonumber
\end{align}
\end{subequations}
where $u_2(\bf z)$ is for notational convenience.  Similarly, we apply Taylor approximation on $u_2(\bf z)$ and obtain a lower bound  $R_{2, k}^{\rm low}$, which has closed-form expression from $\nabla u_2({\bf z}^{\rm loc})$. Then, \eqref{eq_op_AO_Relay_2} can be converted to
\begin{subequations} \label{eq_op_AO_Relay_3}
\begin{align}
& \max
& & \sum_{k=1}^{K}{R_{2,k}} \label{eq_op_AO_Relay_3_obj}  \\
& \text{variables: } & & \{\mathbf{F}_{k}\}, \{R_{2,k}\}, \{{\mathcal S}_{2,k}\}, \{{\mathcal I}_{2,k}\} \nonumber  \\
& \hspace{9pt} \text{s.t.} & & R_{2, k} \leq R_{2, k}^{\rm low}, ~~\eqref{eq_op_AO_Relay_2_cons_supp2}, \label{eq_op_AO_Relay_3_cons1} \\
& & &  \forall k = 1,\ldots,K. \nonumber
\end{align}
\end{subequations}
The solution set of \eqref{eq_op_AO_Relay_3} is a subset of the solution set of \eqref{eq_op_AO_Relay_2}, since the lower bound $R_{2, k}^{\rm low}$ is used in the constraint \eqref{eq_op_AO_Relay_2_cons1}. By relaxing the rank-one constraint in \eqref{eq_op_AO_Relay_2_cons_supp2}, we can solve the resulting SDP problem with CVX. Finally, as in the case in Sec.~\ref{sec_pro_alg_BS}, the randomization procedure can be employed to obtain rank-one solutions for $\{{\mathbf f}_{k}\}$, $k=1,\ldots,K$.

\subsection{IRS Beamforming Optimization} \label{sec_pro_alg_IRS}
Given fixed BS beamforming $\{{\mathbf g}_{k}\}$ and relay beamforming $\{{\mathbf f}_{k}\}$ (respectively obtained from the steps in Sec. \ref{sec_pro_alg_BS} and Sec. \ref{sec_pro_alg_Relay}), the problem of solving the IRS beamforming matrix $\boldsymbol\Theta$ (or equivalently its vector form $\boldsymbol\theta$) in \eqref{eq_op_sum_rate} becomes
\begin{subequations} \label{eq_op_AO_IRS}
\begin{align}
& \max_{\mathbf{\Theta}}
& & \sum_{k=1}^{K}{\log_{2}\left(1 + \gamma_{k}\right)} \label{eq_op_IRS_obj}  \\
& \hspace{9pt} \text{s.t.} & & {\left|\theta_{n}\right|} \leq 1, \forall n = 1,\ldots,N , \label{eq_op_AO_IRS_cons1}  \\
& & & \gamma_{{\rm R},k} \geq \gamma_{\rm R}^{\rm th}, \forall k = 1,\ldots,K. \label{eq_op_AO_IRS_cons2}
\end{align}
\end{subequations}
We adopt similar steps as in Secs. \ref{sec_pro_alg_BS} and \ref{sec_pro_alg_Relay}  to solve \eqref{eq_op_AO_IRS}. To facilitate finding the optimal IRS beamforming matrix $\boldsymbol\Theta$, we re-arrange the order of $\boldsymbol\Theta$ in \eqref{eq_received_signal_I}  and in \eqref{eq_received_signal_II_DF}  respectively to
${y}_{k}^{\rm I}={\boldsymbol {\phi}}^{\rm H} {\mathbf H}_{{\rm B}, {\rm I}, k} {\mathbf x} + w_{k}^{\rm I}$ and $y_{k}^{\rm II}= {\boldsymbol {\phi}}^{\rm H} {\mathbf H}_{{\rm R}, {\rm I}, k} {\mathbf x}_{\rm R} + w_{k}^{\rm II},$
where ${\boldsymbol {\phi}} = {\begin{bmatrix} {\boldsymbol \theta} & 1 \end{bmatrix}}^{\rm H}$, ${\mathbf H}_{{\rm B}, {\rm I}, k} = \begin{bmatrix} {\rm diag}({\mathbf h}_{{\rm IRS}, k}) {\mathbf H}_{{\rm BS}, {\rm IRS}} \\ {\mathbf h}_{{\rm BS}, k} \end{bmatrix}$, and ${\mathbf H}_{{\rm R}, {\rm I}, k} = \begin{bmatrix} {\rm diag}({\mathbf h}_{{\rm IRS}, k}) {\mathbf H}_{{\rm R}, {\rm IRS}} \\ {\mathbf h}_{{\rm R}, k} \end{bmatrix}$.
With the reordering, the slack variables $\{{\mathcal S}_{1,k}\}$, $\{{\mathcal I}_{1,k}\}$, $\{{\mathcal S}_{2,k}\}$, and $\{{\mathcal I}_{2,k}\}$ respectively in \eqref{S_1}, \eqref{I_1}, \eqref{S_2}, and \eqref{I_2} can be rewritten as
\begin{align}
\frac{1}{{\mathcal S}_{1, k}} &= {\left | {\boldsymbol {\phi}}^{\rm H} {\mathbf H}_{{\rm B}, {\rm I}, k} {\mathbf g}_{k} \right |}^{2} \leq {\rm tr}\left(\mathbf{\Phi} {\mathbf H}_{{\rm B}, {\rm I}, k} {\mathbf g}_{k} {\mathbf g}_{k}^{\rm H} {\mathbf H}_{{\rm B}, {\rm I}, k}^{\rm H}\right), \label{S_1_prime}\\
{\mathcal I}_{1,k} & \geq \sum_{j \neq k}  {\rm tr}\left(\mathbf{\Phi} {\mathbf H}_{{\rm B}, {\rm I}, k} {\mathbf g}_{j} {\mathbf g}_{j}^{\rm H} {\mathbf H}_{{\rm B}, {\rm I}, k}^{\rm H}\right) + \sigma_{k}^{2},\label{I_1_prime}\\
\frac{1}{{\mathcal S}_{2,k}} & = {\left | {\boldsymbol {\phi}}^{\rm H} {\mathbf H}_{{\rm R}, {\rm I}, k} {\mathbf f}_{k} \right |}^{2} \leq {\rm tr}(\mathbf{\Phi} {\mathbf H}_{{\rm R}, {\rm I}, k} {\mathbf f}_{k} {\mathbf f}_{k}^{\rm H} {\mathbf H}_{{\rm R}, {\rm I}, k}^{\rm H}),\label{S_2_prime}\\
{\mathcal I}_{2,k} & \geq \sum_{j \neq k} {\rm tr}\left(\mathbf{\Phi} {\mathbf H}_{{\rm R}, {\rm I}, k} {\mathbf f}_{j} {\mathbf f}_{j}^{\rm H} {\mathbf H}_{{\rm R}, {\rm I}, k}^{\rm H}\right) + \sigma_{k}^{2},\label{I_2_prime}
\end{align}
where $\mathbf{\Phi} \triangleq {\boldsymbol {\phi}} {\boldsymbol {\phi}}^{\rm H}$.
Similarly, we can re-arrange the order of $\boldsymbol\Theta$ in \eqref{eq_received_Relay}, by which the slack variables $\{{\mathcal S}_{{\rm R},k}\}$ and $\{{\mathcal I}_{{\rm R},k}\}$ in \eqref{S_R} and \eqref{I_R} can be rewritten in terms of $\mathbf{\Phi}$.

With the above slack variables, we have the following equivalent problem
\begin{subequations} \label{eq_op_AO_IRS_2}
\begin{align}
& \max
& & \sum_{k=1}^{K}{R_{3,k}} \label{eq_op_AO_IRS_2_obj}  \\
& \text{variables: } & & \mathbf{\Phi}, \{R_{3,k},{\mathcal S}_{1,k},{\mathcal I}_{1,k},{\mathcal S}_{2,k},{\mathcal I}_{2,k},{\mathcal S}_{{\rm R},k},{\mathcal I}_{{\rm R},k}\} \nonumber  \\
& \hspace{9pt} \text{s.t.} & & R_{3,k} \leq \log_{2}\left(1 + \frac{1}{{\mathcal S}_{1,k} {\mathcal I}_{1,k}} + \frac{1}{{\mathcal S}_{2,k}{\mathcal I}_{2,k}}\right)\notag\\
& & & \quad\quad\triangleq u_3({\bf z}), \label{eq_op_AO_IRS_2_cons1}\\
& & & v_3({\bf z}_{\rm R})\triangleq\frac{1}{{\mathcal S}_{{\rm R}, k} {\mathcal I}_{{\rm R}, k}} \geq \gamma_{\rm R}^{\rm th}, \label{eq_op_AO_IRS_2_cons2} \\
& & &  \eqref{eq_op_AO_IRS_cons1},\eqref{S_R},\eqref{I_R}, \eqref{S_1_prime}-\eqref{I_2_prime},  \label{eq_op_AO_IRS_2_cons_supp}\\
& & & \mathbf{\Phi} \succeq 0,~~\text{rank}\left(\mathbf{\Phi}\right) = 1, \label{eq_op_AO_IRS_2_cons_rank1}\\
& & &\quad\quad\forall k = 1,\ldots,K,\nonumber
\end{align}
\end{subequations}
where $u_3({\bf z})$ and $v_3({\bf z}_{\rm R})$ respectively defined in \eqref{eq_op_AO_IRS_2_cons1} and \eqref{eq_op_AO_IRS_2_cons2} are for notational convenience. Next, using the first-order Taylor approximation, as in the previous sections, lower bounds of $u_3({\bf z})$ and $v_3({\bf z}_{\rm R})$ can be obtained as $R_{3, k}^{\rm low}$ and $\gamma_{{\rm R}, k}^{\rm low}$, which can be used to convexify the two constraints \eqref{eq_op_AO_IRS_2_cons1} and \eqref{eq_op_AO_IRS_2_cons2}.   
Then, \eqref{eq_op_AO_IRS_2} can be rewritten as
\begin{subequations} \label{eq_op_AO_IRS_3}
\begin{align}
& \max
& & \sum_{k=1}^{K}{R_{3,k}} \label{eq_op_AO_IRS_3_obj}  \\
& \text{variables: } & & \mathbf{\Phi}, \{R_{3,k},{\mathcal S}_{1,k},{\mathcal I}_{1,k},{\mathcal S}_{2,k},{\mathcal I}_{2,k},{\mathcal S}_{{\rm R},k},{\mathcal I}_{{\rm R},k}\} \nonumber  \\
& \hspace{9pt} \text{s.t.} & & R_{3,k} \leq R_{3,k}^{\rm low},~~ \gamma_{{\rm R}, k}^{\rm low} \geq \gamma_{\rm R}^{\rm th}, \eqref{eq_op_AO_IRS_2_cons_supp},\eqref{eq_op_AO_IRS_2_cons_rank1}, \label{eq_op_AO_IRS_3_cons1}\\
& & & \quad\quad \forall k = 1,2, \ldots,K. \notag
\end{align}
\end{subequations}
Relaxing the rank-one constraint  \eqref{eq_op_AO_IRS_2_cons_rank1} in \eqref{eq_op_AO_IRS_3_cons1} converts the problem in \eqref{eq_op_AO_IRS_3} to an SDP, which can then be solved using CVX. Finally, Gaussian randomization procedure can be employed to obtain a rank-one solution. 

\section{Simulation Results and Discussion} \label{sec_num_res}

\subsection{Simulation Settings}

We simulate the topology shown in Fig.~\ref{fig:system_model} with the BS, IRS, and relay located at coordinates $(0,0)$, $(100,50)$, and $(100,-50)$, respectively, and the $K$ users randomly distributed within a circle centered at $(0,200)$ with a radius of $10$. The units of locations and radius are all meters. The small-scale fading of all channels follows $\mathcal{CN} (0,1)$. The parameters for large-scale fading of channels $\kappa (d/d_0)^{-\varrho}$ are set as follows: $d_{0} = 1$ m; $\kappa = 10^{-4}$ for ${\mathbf H}_{{\rm BS}, {\rm R}}$, ${\mathbf h}_{{\rm BS}, k}$, and ${\mathbf h}_{{\rm R}, k}$, and $\kappa = 10^{-0.5}$ for ${\mathbf H}_{{\rm BS}, {\rm IRS}}$, ${\mathbf H}_{{\rm R}, {\rm IRS}}$, and ${\mathbf h}_{{\rm IRS}, k}$; and $\varrho = 2$ for IRS-aided links and relay-aided links (free space), and $\varrho = 3.5$ for the direct link (severe path loss). The power constraints at the BS and relay are $P_{\rm BS}^{\rm max} = 10$ mW and $P_{\rm R}^{\rm max} = 10$ mW, respectively. The proposed coexistence and co-design scheme is compared with the following benchmarks: 1) {\it Relay only:} there is no IRS; only the relay assists transmission with BS and relay beamforming optimized; 
2) {\it Random:} IRS beamforming adopts random phase shifts and fixed amplitudes of one, with BS and relay beamforming optimized; 
and 3) {\it Independent:} IRS assists transmission in the first phase but not in the second phase (turned off), with BS, relay, and IRS beamforming optimized in such case. 
Note that it is termed `Independent' since there is no interaction between IRS and relay in assisting BS transmission.


\subsection{Results and Discussion} \label{results and discussion}

\begin{figure*}[t]
\begin{center}
 \subfigure[]
 {
  \includegraphics[width=0.64\columnwidth]{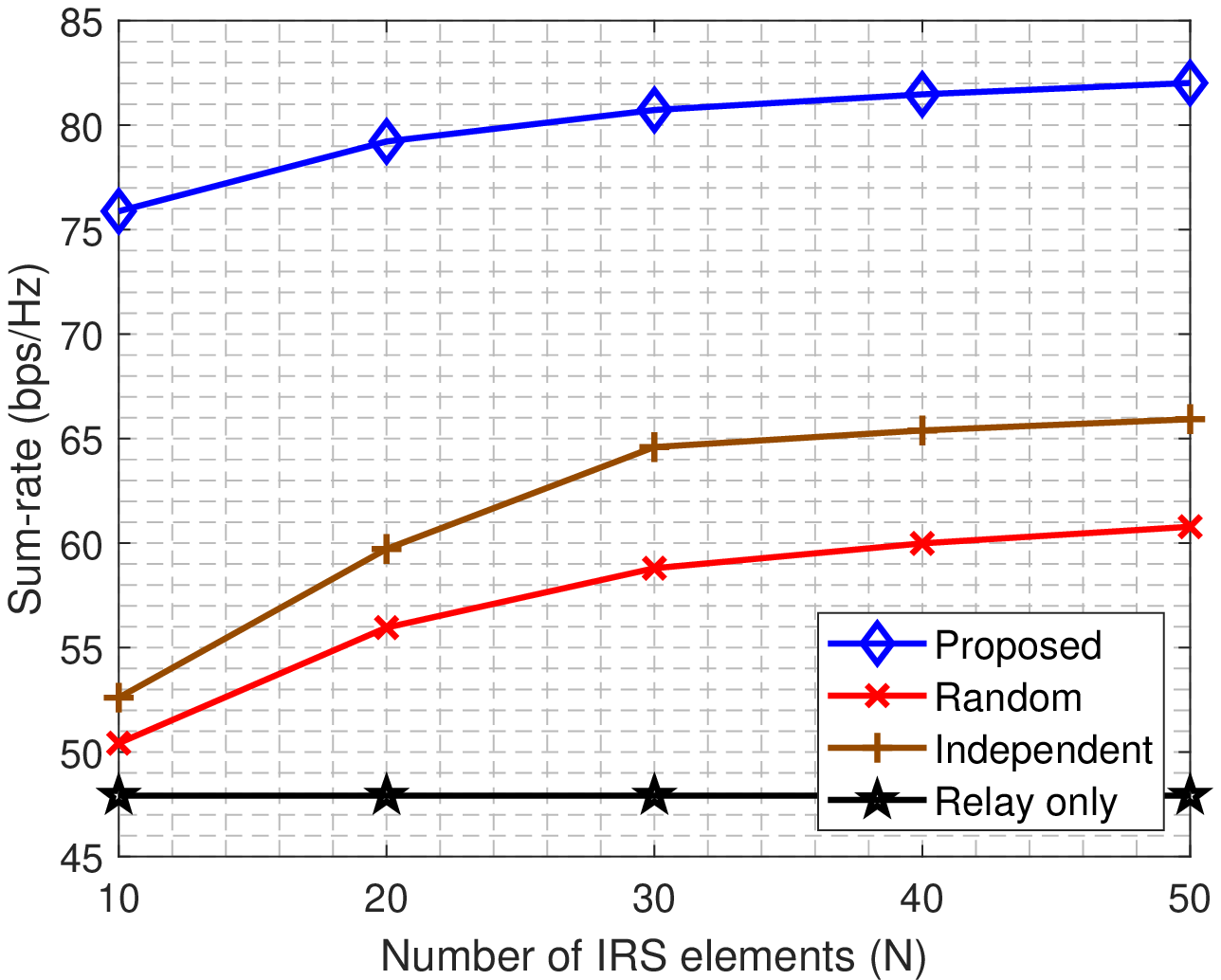}
  \label{fig:simulation_num_elements}
   }
 \subfigure[]
 {
  \includegraphics[width=0.64\columnwidth]{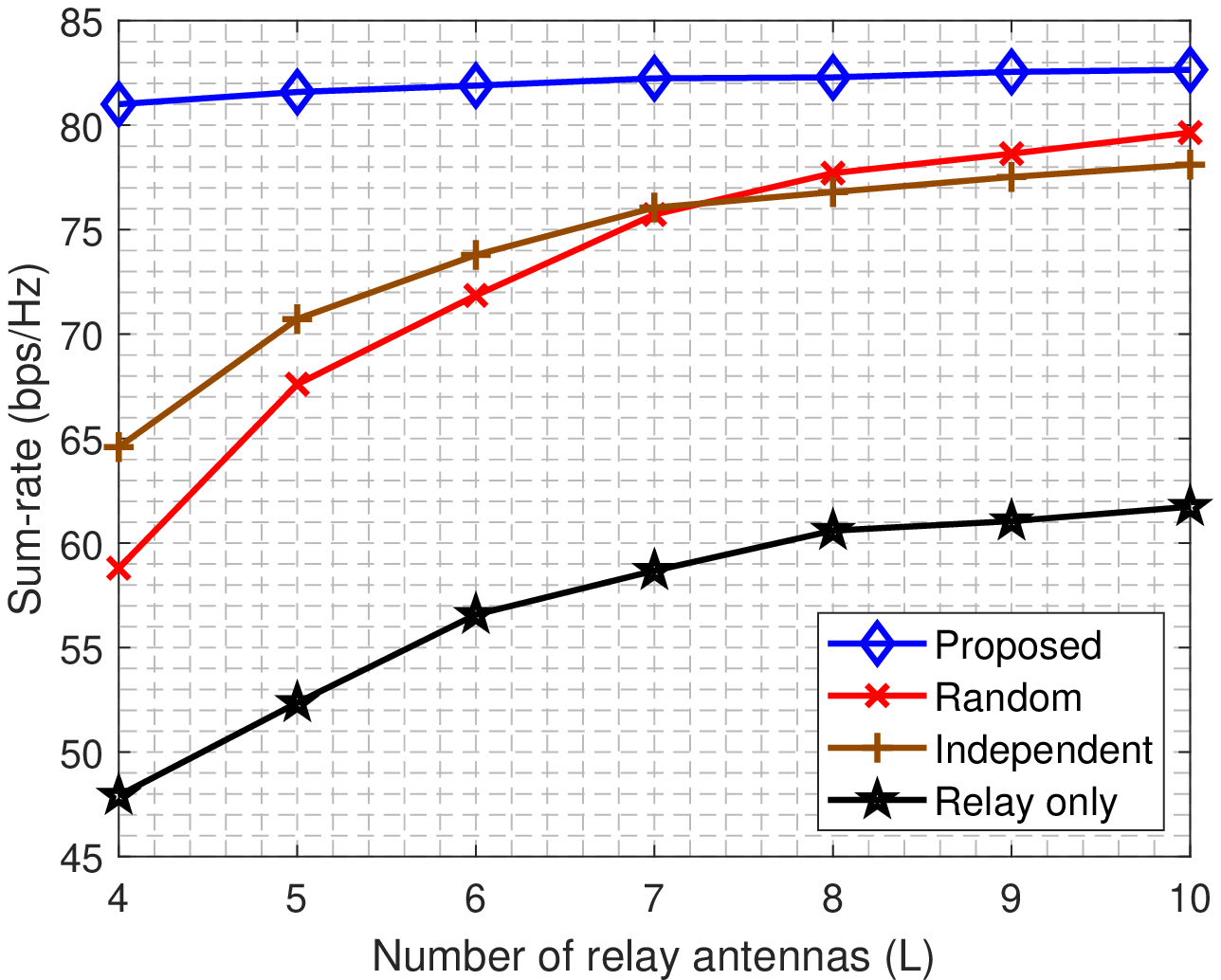}
  \label{fig:simulation_num_DF_antennas}
   }
 \subfigure[]
 {
  \includegraphics[width=0.64\columnwidth]{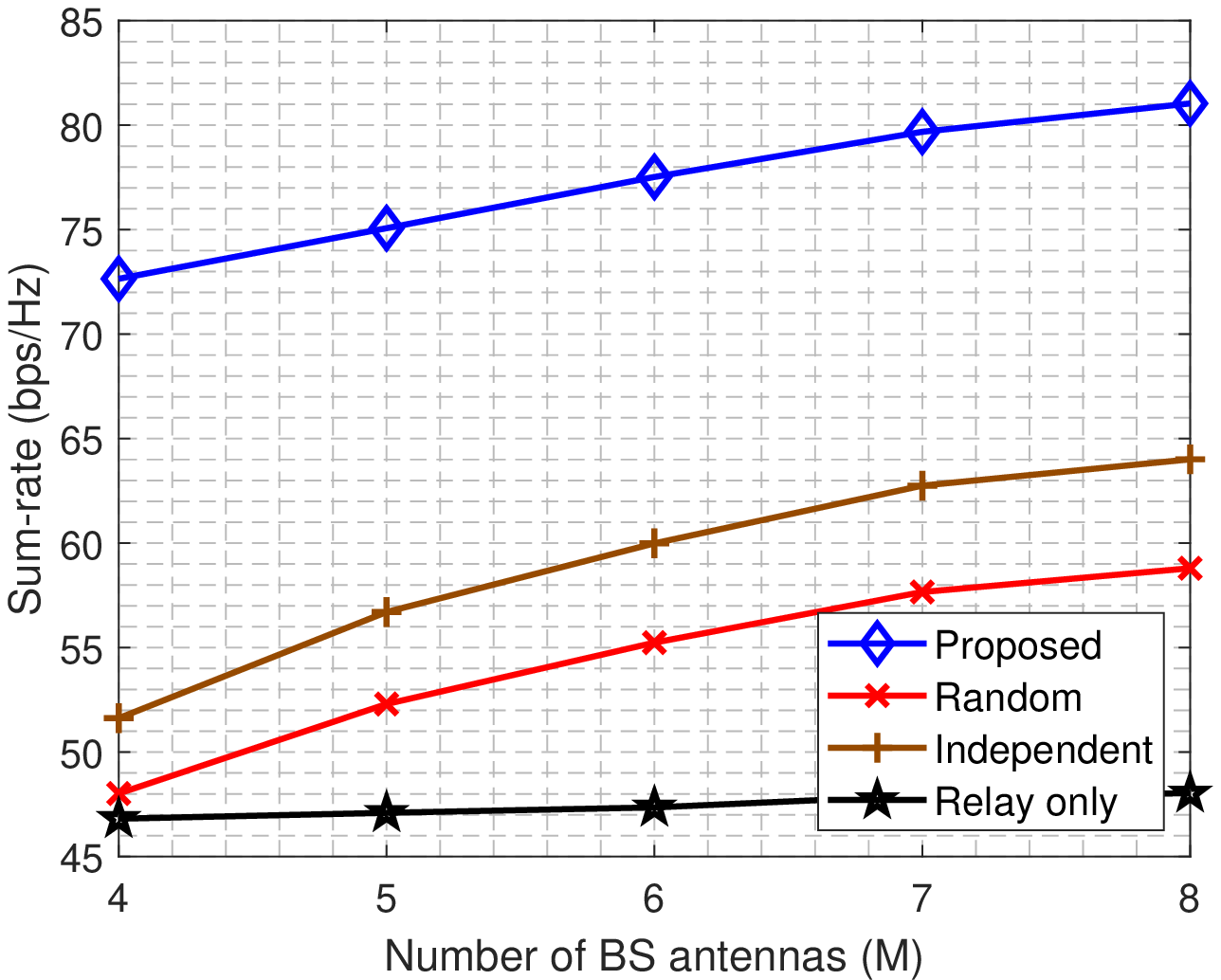}
  \label{fig:simulation_num_BS_antennas}
   }
 \caption{(a)--(c) Sum-rate vs. the number of IRS reflecting elements, relay antennas, and BS antennas, respectively.}
 \label{fig:simulation_results}
\end{center}
\vspace{-0.1in}
\end{figure*}

Fig.~\ref{fig:simulation_num_elements} compares the sum-rate performance vs. the number of IRS reflecting elements $N$, where we set $M = 8$, $L = 4$, and $K = 4$. As can be seen, the sum-rate of all schemes except the Relay only scheme increases as $N$ increases. The proposed scheme outperforms all benchmarks since it fully exploits the coexistence system with optimized beamforming and transmission mechanism. The Independent scheme exhibits performance loss in comparison to the proposed scheme since IRS is turned off in the second phase and thus relaying in the second phase does not acquire performance gain from IRS-enhanced links and IRS beamforming. This suggests that the interaction between IRS and DF relay benefits the system sum-rate. Comparing the Independent and Random schemes, while the Independent scheme only leverages IRS in the first phase, it still outperforms the Random scheme with IRS operating in both phases. This is because, while the Independent scheme achieves a lower SINR in the second phase as compared to the Random scheme, it provides a much higher SINR in the first phase due to judicious IRS beamforming, resulting in an overall higher sum-rate. The Random scheme outperforms the Relay only scheme since IRS, even with random phase shifts, provides additional paths for transmission.

Fig.~\ref{fig:simulation_num_DF_antennas} compares the sum-rate performance vs. the number of relay antennas $L$, where we set $M = 8$, $N = 30$, and $K = 4$. The sum-rate of all schemes increases with $L$ due to a higher degree of freedom in the second phase. The proposed scheme achieves the best sum-rate performance in all antenna configurations, with diminishing gains as $L$ increases. This is because the existence and optimization of IRS with $N=30$ reflecting elements dominates the change in $L$ in terms of the system capacity in the second phase. As a result, increasing the number of relay antennas brings relatively small merits to the proposed scheme. The Independent and Random schemes exhibit a crossing point; specifically, the Random scheme outperforms the Independent scheme in the larger $L$ regime. The reason is that the DF relay with more antennas can better adjust the relay beamforming vectors $\{{\mathbf f}_{k}\}$ in \eqref{eq_SINR_II} for the combined IRS-assisted and direct channels from the relay to end users in the Random scheme. In contrast, since the IRS is turned off in the second phase, a relatively small gain is created for the Independent scheme as $L$ increases. More specifically, the higher SINR in the second phase outweighs the lower SINR in the first phase for the Random scheme, as compared to the Independent scheme.

Fig.~\ref{fig:simulation_num_BS_antennas} plots the sum-rate performance vs. the number of BS antennas $M$, where we set $L = 4$, $N = 30$, and $K = 4$. The performance of all schemes improves as $M$ increases. Increasing $M$ however creates an incremental gain for the Relay only scheme due to the weak direct link from the BS to the end users. Other schemes, in contrast, are supported by the combined IRS-assisted and direct links in the first phase instead of only the direct link, and therefore observe larger gains as $M$ increases. The Independent scheme outperforms the Random scheme because the Independent scheme is enhanced by IRS beamforming in the first phase. Without optimized IRS phase shifts, the Random scheme suffers from performance loss. Especially, in the small $M$ region, the Random scheme achieves only a small performance gain as compared to the Relay only scheme. This suggests that the coexistence of an IRS and a relay does not always significantly benefit communications unless a proper co-design is performed.

\section{Conclusion} \label{sec_conclusion}

We have considered a multiuser MISO system in which an IRS and a DF relay coexist and assist downlink transmission simultaneously. We investigated a co-design problem of BS beamforming, relay beamforming, and IRS beamforming for sum-rate maximization. To tackle the complex nonconvex co-design problem, we proposed an algorithm based on alternating optimization and semidefinite relaxation. Simulation results and discussion on our proposed scheme and other benchmarks were presented. It was shown that our proposed algorithm significantly increases the sum-rate of end users.

\bibliographystyle{IEEEtran}
\bibliography{IEEEabrv,references}

\end{document}